\documentclass[10pt,preprint]{aastex}

\usepackage{epsfig}
\usepackage{graphicx}
\usepackage{multicol}

\slugcomment{Accepted to ApJ}
\shorttitle{Discovery of HE and VHE $\gamma$-Ray Emission from RBS~0413}

\begin{document}

\title{DISCOVERY OF HIGH-ENERGY AND VERY HIGH ENERGY $\gamma$-RAY EMISSION FROM THE BLAZAR RBS~0413}

\author{
E.~Aliu\altaffilmark{1},
S.~Archambault\altaffilmark{2},
T.~Arlen\altaffilmark{3},
T.~Aune\altaffilmark{4},
M.~Beilicke\altaffilmark{5},
W.~Benbow\altaffilmark{6},
M.~B{\"o}ttcher\altaffilmark{7},
A.~Bouvier\altaffilmark{4},
S.~M.~Bradbury\altaffilmark{8},
J.~H.~Buckley\altaffilmark{5},
V.~Bugaev\altaffilmark{5},
K.~Byrum\altaffilmark{9},
A.~Cannon\altaffilmark{10},
A.~Cesarini\altaffilmark{11},
L.~Ciupik\altaffilmark{12},
E.~Collins-Hughes\altaffilmark{10},
M.~P.~Connolly\altaffilmark{11},
P.~Coppi\altaffilmark{13},
W.~Cui\altaffilmark{14},
G.~Decerprit\altaffilmark{9},
R.~Dickherber\altaffilmark{5},
J.~Dumm\altaffilmark{15},
M.~Errando\altaffilmark{1},
A.~Falcone\altaffilmark{16},
Q.~Feng\altaffilmark{14},
J.~P.~Finley\altaffilmark{14},
G.~Finnegan\altaffilmark{17},
L.~Fortson\altaffilmark{15},
A.~Furniss\altaffilmark{4},
N.~Galante\altaffilmark{6},
D.~Gall\altaffilmark{18},
S.~Godambe\altaffilmark{17},
S.~Griffin\altaffilmark{2},
J.~Grube\altaffilmark{12},
G.~Gyuk\altaffilmark{12},
D.~Hanna\altaffilmark{2},
K.~Hawkins\altaffilmark{7},
J.~Holder\altaffilmark{19},
H.~Huan\altaffilmark{20},
G.~Hughes\altaffilmark{21},
T.~B.~Humensky\altaffilmark{22},
P.~Kaaret\altaffilmark{18},
N.~Karlsson\altaffilmark{15},
M.~Kertzman\altaffilmark{23},
Y.~Khassen\altaffilmark{10},
D.~Kieda\altaffilmark{17},
H.~Krawczynski\altaffilmark{5},
F.~Krennrich\altaffilmark{24},
M.~J.~Lang\altaffilmark{11},
K.~Lee\altaffilmark{5},
A.~S~Madhavan\altaffilmark{24},
G.~Maier\altaffilmark{21},
P.~Majumdar\altaffilmark{3},
S.~McArthur\altaffilmark{5},
A.~McCann\altaffilmark{2},
P.~Moriarty\altaffilmark{25},
R.~Mukherjee\altaffilmark{1},
R.~A.~Ong\altaffilmark{3},
M.~Orr\altaffilmark{24},
A.~N.~Otte\altaffilmark{4},
N.~Palma\altaffilmark{7},
N.~Park\altaffilmark{20},
J.~S.~Perkins\altaffilmark{26,27},
A.~Pichel\altaffilmark{28},
M.~Pohl\altaffilmark{21,29},
H.~Prokoph\altaffilmark{21},
J.~Quinn\altaffilmark{10},
K.~Ragan\altaffilmark{2},
L.~C.~Reyes\altaffilmark{30},
P.~T.~Reynolds\altaffilmark{31},
E.~Roache\altaffilmark{6},
H.~J.~Rose\altaffilmark{8},
J.~Ruppel\altaffilmark{21,29},
D.~B.~Saxon\altaffilmark{19},
M.~Schroedter\altaffilmark{6},
G.~H.~Sembroski\altaffilmark{14},
G.~D.~\c{S}ent\"{u}rk\altaffilmark{22},
A.~W.~Smith\altaffilmark{17},
D.~Staszak\altaffilmark{2},
I.~Telezhinsky\altaffilmark{21,29},
G.~Te\v{s}i\'{c}\altaffilmark{2},
M.~Theiling\altaffilmark{14},
S.~Thibadeau\altaffilmark{5},
K.~Tsurusaki\altaffilmark{18},
A.~Varlotta\altaffilmark{14},
M.~Vivier\altaffilmark{19},
S.~P.~Wakely\altaffilmark{20},
J.~E.~Ward\altaffilmark{10},
T.~C.~Weekes\altaffilmark{6},
A.~Weinstein\altaffilmark{24},
T.~Weisgarber\altaffilmark{20},
D.~A.~Williams\altaffilmark{4},
B.~Zitzer\altaffilmark{14}
}

\author{
P.~Fortin\altaffilmark{32}, 
D.~Horan\altaffilmark{32}
}
\altaffiltext{1}{Department of Physics and Astronomy, Barnard College, Columbia University, NY 10027, USA}
\altaffiltext{2}{Physics Department, McGill University, Montreal, QC H3A 2T8, Canada}
\altaffiltext{3}{Department of Physics and Astronomy, University of California, Los Angeles, CA 90095, USA}
\altaffiltext{4}{Santa Cruz Institute for Particle Physics and Department of Physics, University of California, Santa Cruz, CA 95064, USA}
\altaffiltext{5}{Department of Physics, Washington University, St. Louis, MO 63130, USA}
\altaffiltext{6}{Fred Lawrence Whipple Observatory, Harvard-Smithsonian Center for Astrophysics, Amado, AZ 85645, USA}
\altaffiltext{7}{Astrophysical Institute, Department of Physics and Astronomy, Ohio University, Athens, OH 45701}
\altaffiltext{8}{School of Physics and Astronomy, University of Leeds, Leeds, LS2 9JT, UK}
\altaffiltext{9}{Argonne National Laboratory, 9700 S. Cass Avenue, Argonne, IL 60439, USA}
\altaffiltext{10}{School of Physics, University College Dublin, Belfield, Dublin 4, Ireland}
\altaffiltext{11}{School of Physics, National University of Ireland Galway, University Road, Galway, Ireland}
\altaffiltext{12}{Astronomy Department, Adler Planetarium and Astronomy Museum, Chicago, IL 60605, USA}
\altaffiltext{13}{Department of Astronomy, Yale University, P.O. Box 208101, New Haven CT, 06520-8101 USA}
\altaffiltext{14}{Department of Physics, Purdue University, West Lafayette, IN 47907, USA }
\altaffiltext{15}{School of Physics and Astronomy, University of Minnesota, Minneapolis, MN 55455, USA}
\altaffiltext{16}{Department of Astronomy and Astrophysics, 525 Davey Lab, Pennsylvania State University, University Park, PA 16802, USA}
\altaffiltext{17}{Department of Physics and Astronomy, University of Utah, Salt Lake City, UT 84112, USA}
\altaffiltext{18}{Department of Physics and Astronomy, University of Iowa, Van Allen Hall, Iowa City, IA 52242, USA}
\altaffiltext{19}{Department of Physics and Astronomy and the Bartol Research Institute, University of Delaware, Newark, DE 19716, USA}
\altaffiltext{20}{Enrico Fermi Institute, University of Chicago, Chicago, IL 60637, USA}
\altaffiltext{21}{DESY, Platanenallee 6, 15738 Zeuthen, Germany}
\altaffiltext{22}{Physics Department, Columbia University, New York, NY 10027, USA; gunessenturk@gmail.com}
\altaffiltext{23}{Department of Physics and Astronomy, DePauw University, Greencastle, IN 46135-0037, USA}
\altaffiltext{24}{Department of Physics and Astronomy, Iowa State University, Ames, IA 50011, USA}
\altaffiltext{25}{Department of Life and Physical Sciences, Galway-Mayo Institute of Technology, Dublin Road, Galway, Ireland}
\altaffiltext{26}{CRESST and Astroparticle Physics Laboratory NASA/GSFC, Greenbelt, MD 20771, USA.}
\altaffiltext{27}{University of Maryland, Baltimore County, 1000 Hilltop Circle, Baltimore, MD 21250, USA.}
\altaffiltext{28}{Instituto de Astronomia y Fisica del Espacio, Casilla de Correo 67 - Sucursal 28, (C1428ZAA) Ciudad Autónoma de Buenos Aires, Argentina}
\altaffiltext{29}{Institut f\"ur Physik und Astronomie, Universit\"at Potsdam, 14476 Potsdam-Golm,Germany}
\altaffiltext{30}{Physics Department, California Polytechnic State University, San Luis Obispo, CA 94307, USA}
\altaffiltext{31}{Department of Applied Physics and Instrumentation, Cork Institute of Technology, Bishopstown, Cork, Ireland}


\altaffiltext{32}{Laboratoire Leprince-Ringuet, \'Ecole polytechnique, CNRS/IN2P3, Palaiseau, France; fortin@llr.in2p3.fr, deirdre@llr.in2p3.fr}

\begin{abstract}
We report on the discovery of high-energy (HE; $E\,>\,0.1$\,GeV) and very high-energy (VHE; $E\,>\,100$\,GeV) $\gamma$-ray emission from the high-frequency-peaked BL Lac object RBS\,0413. 
VERITAS, a ground-based $\gamma$-ray observatory, detected VHE $\gamma$ rays from RBS\,0413 with a statistical significance of 5.5 standard deviations ($\sigma$) and a $\gamma$-ray flux of $(1.5\pm0.6_\mathrm{stat}\pm0.7_\mathrm{syst})\times10^{-8}$~$\textrm{photons}$~$\textrm{m}^{-2}$~$\textrm{s}^{-1}$ ($\sim1\%$ of the Crab Nebula flux) above 250\,GeV. 
The observed spectrum can be described by a power law with a photon index of $3.18\pm0.68_\mathrm{stat}\pm0.30_\mathrm{syst}$. Contemporaneous observations with the Large Area Telescope (LAT) on the \emph{Fermi} Gamma-ray Space Telescope detected HE $\gamma$ rays from RBS\,0413 with a statistical significance of more than $9\sigma$, a power-law photon index of $1.57\pm{0.12_\mathrm{stat}}^{+0.11}_{-0.12\mathrm{sys}}$ and a $\gamma$-ray flux between 300\,MeV and 300\,GeV of ($1.64\pm{0.43_\mathrm{stat}}^{+0.31}_{-0.22\mathrm{sys}}$) $\times$ 10$^{-5}$ photons m$^{-2}$ s$^{-1}$. 
We present the results from \emph{Fermi}-LAT and VERITAS, including a spectral energy distribution modeling of the $\gamma$-ray, quasi-simultaneous X-ray (\emph{Swift}-XRT), ultraviolet (\emph{Swift}-UVOT) and $R$-band optical (MDM) data. 
We find that, if conditions close to equipartition are required, both the combined synchrotron self-Compton/external-Compton and
the lepto-hadronic models are preferred over a pure synchrotron self-Compton model.

\end{abstract}


\clearpage


\section{Introduction}

Blazars are active galactic nuclei that have their jet axis oriented at a small angle with respect to the observer~\citep{urry95}. 
They are observationally classified as either flat-spectrum radio quasars (FSRQ) or BL Lacertae (BL Lac) objects according to the broad line emission in their optical spectra. 
Recent studies interpreting the differing spectral high-energy (HE) $\gamma$-ray properties of the FSRQs and BL Lacs based on physical mechanisms can be found in e.g.,~\citet{ghisellini09}.
Blazars are known to emit non-thermal radiation characterized by a double-peaked spectral energy distribution (SED). 
The low-energy component, generally covering radio to UV/X-ray bands, is usually explained as due to synchrotron emission from relativistic electrons in the blazar jet. 
The origin of the HE component, occurring in the X-ray to $\gamma$-ray regime, is still not completely resolved and could be due to emission from a relativistic particle beam consisting of leptons and / or hadrons. 
In leptonic models, very high energy (VHE) photons are produced by inverse-Compton (IC) scattering of low-energy photons off the synchrotron-emitting electrons. 
The soft seed photons for the IC process can be the synchrotron photons (synchrotron self-Compton, SSC, e.g.,~\citet{maraschi92}), or they may originate from ambient radiation (external-Compton, EC, e.g.,~\citet{dermer93}).
Hadronic models include synchrotron emission from protons (e.g.,~\citet{aharonian00}) and $\pi^\mathrm{0}$-decay from hadronic interactions with subsequent electromagnetic cascades (e.g.,~\citet{mucke03}).
\par
RBS\,0413 was discovered in the X-ray band (1E\,0317.0+1834) during the Einstein Medium Sensitivity Survey and was optically identified as a BL Lac~\citep{gioia84}. 
The object was also detected as a radio emitter with the Very Large Array of the National Radio Astronomy Observatory~\citep{stocke90}. 
It exhibits significant and variable optical polarization~\citep{stocke85}.
Having a ``featureless'' optical spectrum~\citep{stocke89} and an estimated synchrotron peak frequency log$(\nu_\mathrm{peak}/\mathrm{Hz})=16.99$~\citep{nieppola06}, RBS\,0413 is classified as a high-frequency-peaked BL Lac object (HBL,~\citet{padovani95}). 
It is located at a redshift of 0.190~\citep{gioia84, stocke85}.
\par
The MAGIC Collaboration observed RBS\,0413 in 2004 December--2005 February for a livetime of 6.9 hr and reported a VHE flux upper limit of  $4.2\times10^{-12}~\textrm{erg}~\textrm{cm}^{-2}~\textrm{s}^{-1}$, at 200\,GeV, assuming a power-law spectrum with a photon index of 3.0~\citep{albert08}. 
VERITAS observed the source in the 2008--2009 season and obtained a marginal significance of $\sim3\sigma$.
In 2009, \emph{Fermi}-Large Area Telescope (LAT) detected HE emission from the direction of RBS\,0413~\citep{abdo2010b}, triggering new VERITAS observations. These new observations, combined with the previous data, resulted in the detection of RBS\,0413 as a VHE $\gamma$-ray emitter in 2009 October~\citep{ong2009}. 
\par

\section{VERITAS Observations and Analysis Results}\label{veritas}
VERITAS is a ground-based $\gamma$-ray observatory sensitive to $\gamma$ rays with energy between 100\,GeV and 30\,TeV. 
Located at the Fred Lawrence Whipple Observatory (FLWO) near Amado, southern Arizona, USA (1.3~km above sea level, N $31^{\circ}$ $40'$, W $110^{\circ}$ $57'$), the array consists of four imaging atmospheric Cherenkov telescopes, each having a diameter of 12\,m and a field of view of $3^{\circ}.5$~\citep{holder08}. 
During the 2009 annual shutdown (July\,-\,August) one of the telescopes was relocated to give a more symmetrical array layout. 
In addition, a new mirror-alignment system applied in 2009~Spring contributed to an improvement in the point-spread function (PSF), with a decrease of $25\%-30\%$ in the 80\% containment radius~\citep{mccann2010}. 
As a consequence, the sensitivity showed a significant improvement, and the observation time required for a $1\%$ Crab Nebula detection dropped from $\sim48$ hr to less than 30 hr. 
For observations at $70^{\circ}$ elevation, the energy resolution is $15\%-20\%$, and the angular resolution, defined as the $68\%$ containment radius, is less than $0.1^{\circ}$~\citep{perkins09}.
\par
VERITAS observed RBS\,0413 for $48$ hours in total, using \emph{wobble} mode~\citep{aharonian01}, with north, south, east and west wobble positions.
After discarding observing runs compromised by bad weather, and a small number affected by hardware problems, 26 hours remained for analysis. One third of these data were obtained with the old array configuration (Sep~2008 - Feb~2009, MJD 54732--54883) and the rest with the new array (Sep~2009 - Jan~2010, MJD 55092--55485).
Approximately $3$ hours of data with the old array were taken under weak moonlight, which leads to a higher energy threshold for those observations. 
The source elevation in the data set ranges from $57^{\circ}$ to $79^{\circ}$, with an average of $\sim70^{\circ}$.
Data analysis steps consist of calibration, image parameterization~\citep{hillas85}, event reconstruction, background rejection and signal extraction as described in~\citet{daniel08}. 
For signal extraction, a $\theta^{2}$ cut~\citep{daniel08} of 0.0169, optimized for a point source of 1\% strength of the Crab Nebula, was used. 
\par
\begin{figure}
\centering
\includegraphics[width=0.65\textwidth]{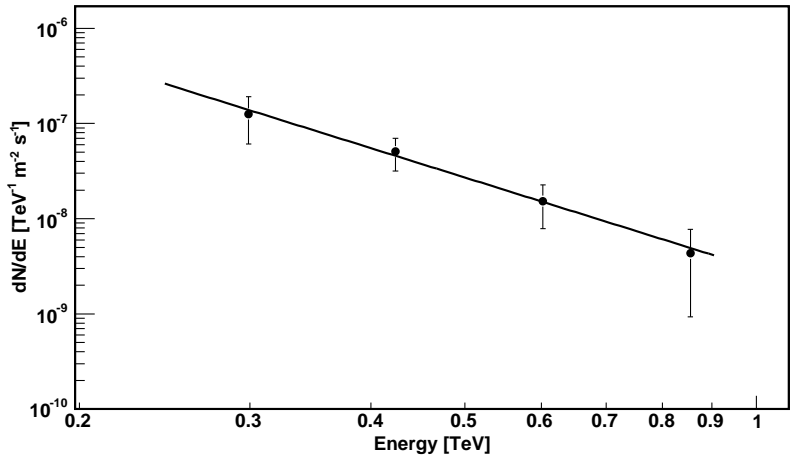}
\caption{{\small VERITAS measured photon spectrum of RBS\,0413. See the text for the parameters of the power-law fit shown.}}
\label{fig:verSpectrum}
\end{figure}

RBS\,0413 is a weak source in the VHE regime.
Using a ``reflected-region'' background estimation~\citep{aharonian01}, an excess of 180 events and a significance of $5.5\sigma$ are obtained, for the source location at $\textrm{RA}=03^{\mathrm{h}}19^{\mathrm{m}}47^{\mathrm{s}}\pm4^{\mathrm{s}}_\mathrm{stat}\pm7^{\mathrm{s}}_\mathrm{syst}$ and $\textrm{decl.}=18^{\circ}45'.7\pm1'.0_\mathrm{stat}\pm1'.8_\mathrm{syst}$ (J2000 coordinates). 
The VERITAS signal is consistent with a point source, and we name the object VER\,J0319+187.
The energy distribution of $\gamma$-ray events extends from $\sim250$\,GeV to $\sim1.0$\,TeV (see Table~\ref{tab:verSpectrum} for a list of spectral data points) and is well described by a power-law function, $dN/dE=F_{0}E^{-\Gamma}$.
The best fit is obtained with photon index $\Gamma=3.18\pm0.68_\mathrm{stat}\pm0.30_\mathrm{syst}$ and flux normalization $F_0=(1.38\pm0.52_\mathrm{stat}\pm0.60_\mathrm{syst})\times10^{-7}$~$\textrm{TeV}^{-1}$~$\textrm{m}^{-2}$~$\textrm{s}^{-1}$ at 0.3 TeV, with a value of $\chi^2$ per degree of freedom ($\chi^2$/dof) of 0.14/2 (see Figure~\ref{fig:verSpectrum}).
\par
The integral flux above 250\,GeV is $(1.5\pm0.6_\mathrm{stat}\pm0.7_\mathrm{syst})\times10^{-8}$~$\textrm{m}^{-2}$~$\textrm{s}^{-1}$, corresponding to a flux level of approximately $1\%$ the flux of the Crab Nebula.
No significant flux variability is detected (see Figure~\ref{fig:lightCurves} top panel and caption for details of the light-curve analysis). 
An upper limit (99\% confidence level) on the fractional variability amplitude ($F_\mathrm{var}$,~\citet{vaughan2003}) yields $F_\mathrm{var} < 3.2$.

\begin{table}
\centering
\begin{small}
\begin{tabular}{ccccc}
\hline\hline
Energy & Flux & Significance\\

(TeV)& (m$^{-2}$\,s$^{-1}$\,TeV$^{-1}$) & ($\sigma$) & \\
\hline
0.30 	& $(1.3\pm0.7)\times10^{-7}$ & 2.1\\
0.42 	& $(5.1\pm1.9)\times10^{-8}$ & 3.0\\
0.60 	& $(1.5\pm0.7)\times10^{-8}$ & 2.3\\
0.85 	& $(4.3\pm3.4)\times10^{-9}$ & 1.2\\

\hline\hline
\end{tabular}
\end{small}
\caption{{\small Differential flux measurements of RBS\,0413 above 250\,GeV with VERITAS. The first column shows the mean energies, weighted by the spectral index. The errors are statistical only.}}
\label{tab:verSpectrum}
\end{table}

\section{\textit{Fermi} Observations and Analysis Results}
The LAT aboard the \emph{Fermi} Gamma-ray Space Telescope is a pair-conversion $\gamma$-ray detector sensitive to photons in the energy range from below 20\,MeV to more than 300\,GeV \citep{atwood09}. The present analysis includes the data taken between 2008 August 4 and 2011 January 4 (MJD 54682--55565), which covers the entire VERITAS observation interval. 
Events from the Pass 6 \emph{diffuse} class with energy between 300\,MeV and 300\,GeV, with zenith angle $<100^{\circ}$, and from a square region of side $20^{\circ}$ centered on RBS\,0413, were selected for this analysis. 
The cut at 300\,MeV was used to minimize larger systematic errors at lower energies. 
The time intervals when the source was close to the Sun (MJD 54954-54974 and 55320-55339) were excluded.
The data were analyzed with the LAT Science Tools version v9r20p0\footnote{http://fermi.gsfc.nasa.gov/ssc/data/analysis/scitools/overview.html} and the post-launch instrument-response functions P6\_V11\_DIFFUSE. 
The binned maximum-likelihood tools were used for significance and flux calculation \citep{cash79,mattox96}. 
Sources from the 1FGL catalog~\citep{abdo2010a} located within a square region of side $24^{\circ}$ centered on RBS\,0413 were included in the model of the region. 
The background model includes the standard Galactic and isotropic diffuse emission components\footnote{http://fermi.gsfc.nasa.gov/ssc/data/access/lat/BackgroundModels.html}.
\par
\begin{figure}
\centering
\includegraphics[width=1.00\textwidth]{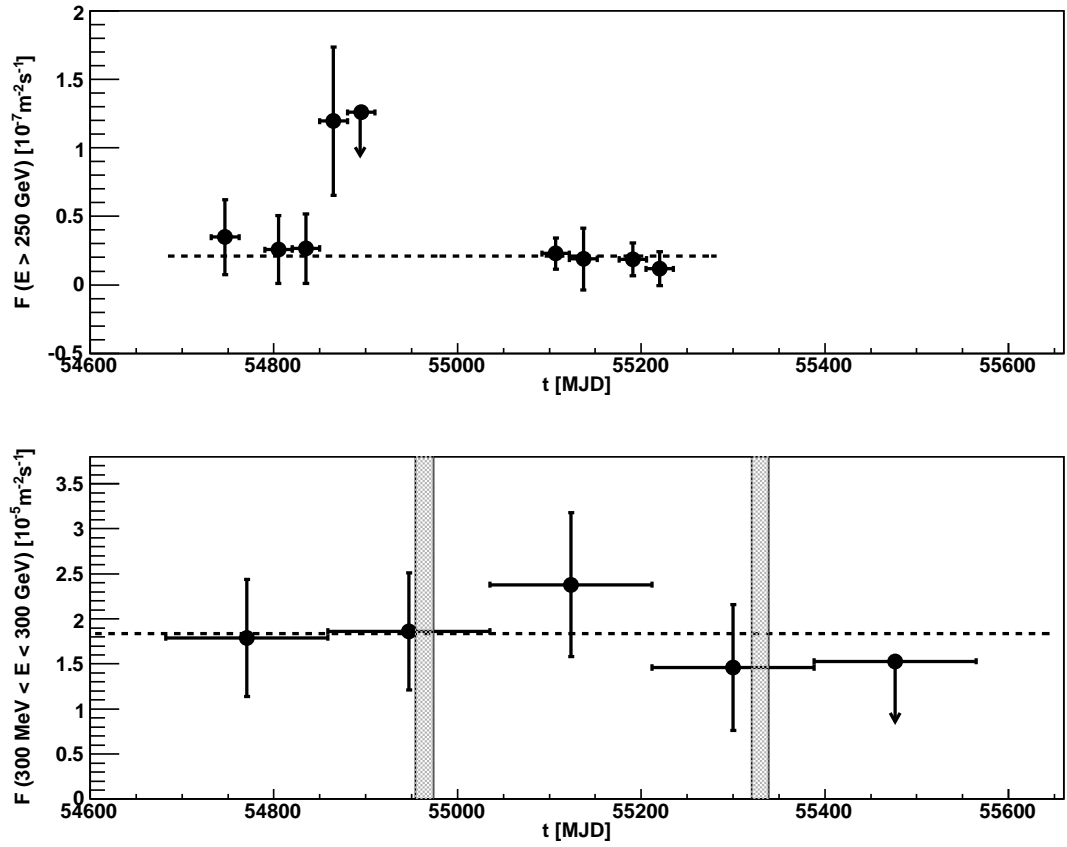}
\caption{{\small Top: 30-day light curve for the VERITAS data. A fit with a constant function gives a $\chi^2$/dof value of 14/8, corresponding to a fit probability of $8\%$, consistent with the hypothesis of a constant flux. The negative flux point corresponding to the upper limit point in the light curve was included in the fit.
Bottom: the light curve for the \emph{Fermi} data using $\sim6$ month wide time bins.
The shaded areas represent the time intervals that were excluded to avoid solar contamination. In both graphs, the dashed lines represent the constant fit function.}}
\label{fig:lightCurves}
\end{figure}

A point source positionally consistent with RBS\,0413 is detected with a significance of more than $9\sigma$ (test statistic, TS=89; see~\citet{mattox96}). 
The photon energy spectrum is best described by a power-law function. 
Replacing the power-law model with a log-parabola model does not significantly improve the likelihood fit. 
The time-averaged integral flux is $I(300\,\mathrm{MeV}<E<300\,\mathrm{GeV})=(1.64{\pm0.43_\mathrm{stat}}^{+0.31}_{-0.22_\mathrm{sys}})\times10^{-5}$~$\textrm{m}^{-2}$~$\textrm{s}^{-1}$, and the spectral index is $1.57{\pm 0.12_\mathrm{stat}}^{+0.11}_{-0.12_\mathrm{sys}}$. 
The spectral points were calculated using the procedure presented in \citet{abdo2010b} (see Table~\ref{tab:fermiSpectrum}). 
In the energy range 100--300\,GeV, no detection was obtained ($\mathrm{TS}<9$) and an upper limit at the 95\% confidence level was derived. 
Figure~\ref{fig:lightCurves} (bottom panel) shows the \emph{Fermi} light curve with $\sim6$ month wide time bins. 
The upper limit point in the last time bin has 95\% confidence level. 
Comparing the likelihood of a model in which the flux in each time bin is free to vary to one where it is assumed to be constant yields a null hypothesis probability of 11\%; thus we find no evidence for variability. 
Details on the methodology can be found in \citet{abdo2011}.
Using the same method as in Section~\ref{veritas} and including the flux value returned by the likelihood fit in the last bin, we estimate $F_\mathrm{var} < 3.2$ with 99\% confidence level.

\begin{table}
\centering
\begin{small}
\begin{tabular}{ccccc}
\hline\hline
Energy & Flux \\

(GeV)& (m$^{-2}$\,s$^{-1}$\,TeV$^{-1}$) \\
\hline
0.55 	& $<2.2\times10^{-2}$	\\
1.73 	& $(2.6\pm0.9)\times10^{-3}$ \\
5.48 	& $(3.9\pm1.2)\times10^{-4}$ \\
17.3 	& $(4.6\pm2.1)\times10^{-5}$ \\
54.8 	& $(7.7\pm4.6)\times10^{-6}$ \\
173 	& $<4.7\times10^{-6}$ \\

\hline\hline
\end{tabular}
\end{small}
\caption{{\small Differential flux measurements of RBS\,0413 with the \emph{Fermi}-LAT. The energies correspond to the bin centers. The errors are statistical only.}}
\label{tab:fermiSpectrum}
\end{table}


\section{\emph{Swift} Observations}

The VERITAS detection triggered a \emph{Swift} \citep{Gehrels04} target-of-opportunity observation of RBS\,0413 on 2009 November 11, with a total exposure of 2.4 ks. All \emph{Swift}-XRT data were reduced using the standard \emph{Swift} analysis pipeline described in~\citet{Burrows05} using the \emph{HEAsoft} 6.8 package. Event files were calibrated and cleaned following the standard filtering criteria using the \emph{xrtpipeline} task and applying calibration files current to 2010 March. All data were taken in photon-counting mode over the energy range 0.3--10 keV. Due to the moderate count rate of 0.3 counts s$^{-1}$, the data are not affected by photon pile-up in the core of the PSF, and partial masking of the source is not necessary. Source events were extracted from a circular region with a radius of 30 pixels ($70\arcsec.8$) centered on the source, and background events were extracted from a 40 pixel radius circle in a source-free region. Ancillary response files were generated using the \emph{xrtmkarf} task, with corrections applied for the PSF losses and CCD defects. The latest response matrix from the XRT calibration files was applied. The extracted XRT energy spectrum was rebinned to contain a minimum of 20 counts in each bin.
\par
An absorbed power-law model, including the \emph{phabs}\footnote{The \emph{phabs} tool applies absorption using photoelectric cross-sections.} model for photoelectric absorption, was fitted to the \emph{Swift}-XRT photon spectrum. 
The cross-sections and abundances used the standard Xspec v12.5 values, as given in the Xspec Analysis Manual\footnote{http://heasarc.nasa.gov/docs/software/lheasoft/xanadu/xspec/XspecManual.pdf}.
Using a fixed Galactic hydrogen column density, $N_{\rm{H}} = 8.91 \times 10^{20}$ cm$^{-2}$ \citep{Kalberla05}, the best-fit model yields a $\chi^{2}$/dof value of 25.9/26.
Over the energy range 0.3--10 keV, the best-fit photon index is $\Gamma = 2.22 \pm 0.07$, and the normalization at 1 keV is ($33.1 \pm 2.2$) keV$^{-1}$ m$^{-2}$ s$^{-1}$. 
The unabsorbed integral flux is $F$(0.3-10\,keV) = ($1.69\pm0.12)\times10^{-11}$~$\mathrm{erg}$~$\mathrm{cm}^{-2}$~$\mathrm{s}^{-1}$ in the range 0.3--10\,keV. The absorbed integral flux in the range 2--10\,keV is $F$(2-10\,keV) = ($5.81\pm0.55)\times10^{-12}$~$\mathrm{erg}$~$\mathrm{cm}^{-2}$~$\mathrm{s}^{-1}$.
No flux variability is evident over the 2.4 ks exposure.
\par
UVOT observations were taken in the photometric band \emph{UVM2} (2246\,\AA) \citep{Poole08}. The \emph{uvotsource} tool was used to extract counts, correct for coincidence losses, apply background subtraction, and calculate the source flux. The standard $5\arcsec$ radius source aperture was used, with a $20\arcsec$ background region. The source fluxes were dereddened using the procedure in~\citet{Roming09}. The measured flux is ($2.75 \pm 0.11$) $\times 10^{-12}$ $\textrm{erg}$~$\textrm{cm}^{-2}$~$\textrm{s}^{-1}$.

\section{MDM Observations}
The $R$-band optical data were taken with the 1.3\,m McGraw-Hill telescope at the MDM observatory on Kitt Peak, Arizona, between 2009 December 10 and 13 . 
All frames were bias corrected and flat fielded using standard routines in IRAF\footnote{IRAF is distributed by the National Optical Astronomy Observatory, which is operated by the Association of Universities for Research in Astronomy (AURA) under cooperative agreement with the National Science Foundation.}~\citep{Barnes93}, and instrumental magnitudes of RBS\,0413 and six comparison stars in the same field of view were extracted using DAOPHOT~\citep{massey92} within IRAF. 
Physical magnitudes were computed using the physical $R$-band magnitudes of the six comparison stars from the NOMAD catalog~\citep{nomad}, assuming that the magnitudes quoted in that catalog are exact, then were corrected for Galactic extinction using extinction coefficients calculated following~\citet{schlegel98}, taken from NED\footnote{http://ned.ipac.caltech.edu/}, and converted into $\nu\mathrm{F}_\mathrm{\nu}$ fluxes. 
The flux shows variations of up to $\sim30\%$ from day to day, with an average of $(2.47\pm0.02)\times10^{-12}$~$\mathrm{erg}$~$\mathrm{cm}^{-2}$~$\mathrm{s}^{-1}$.
In the case of RBS 0413, the host galaxy is expected to make a substantial contribution to the observed $R$-band flux. 
We have taken this into account in our SED modeling by adding a phenomenological host galaxy SED to our model.

\section{Modeling and Discussion}\label{modeling}
\begin{figure}
\centering
\includegraphics[width=0.90\textwidth]{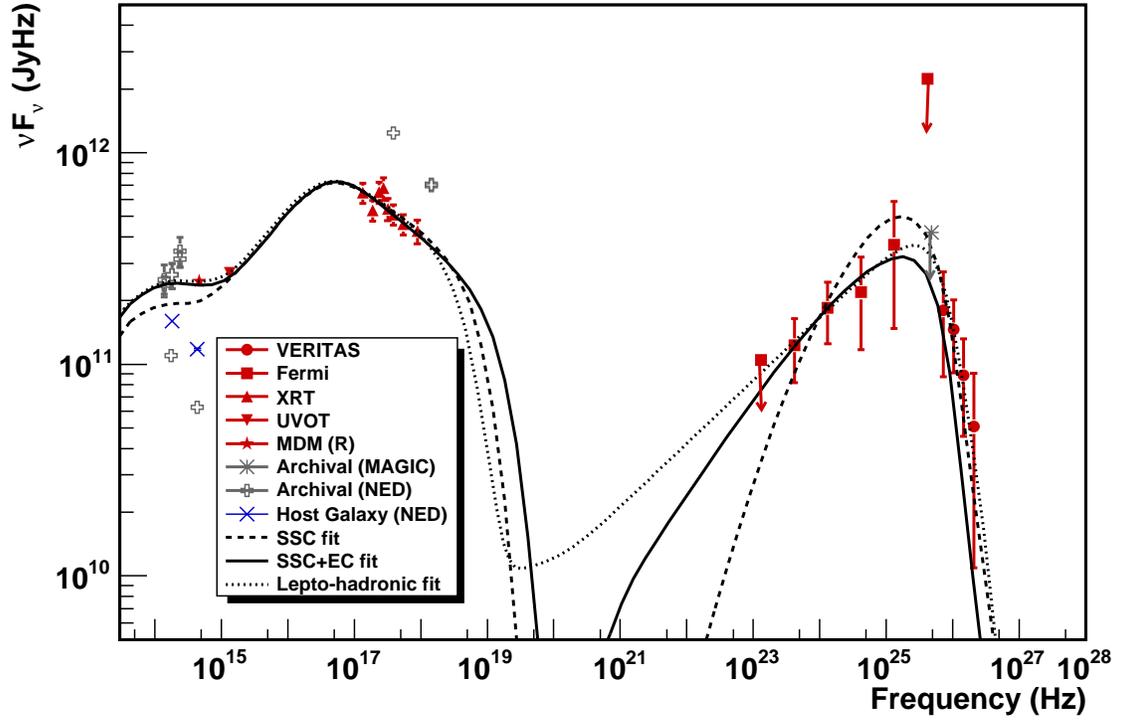}
\caption{{\small RBS\,0413 spectral energy distribution. Absorption in the VHE region due to the EBL is taken into account in the fits using the model of \citet{finke10}. The models are described in detail in the text.}}
\label{fig:SED}
\end{figure}

The non-thermal continuum of RBS\,0413 exhibits a double-peaked shape, as is typical for blazars.
In this study, we applied three different time-independent models to the observed SED, using the contemporaneous X-ray, UV and optical ($R$-band) data to complement the \emph{Fermi}-LAT and VERITAS observations (see Figure~\ref{fig:SED}). 
It should be noted that these observations were not strictly simultaneous.
For all of the models, the emission region was assumed to be a spherical blob of size $R_\mathrm{b}$, moving within the jet with a bulk Lorentz factor $\Gamma$. 
$R_\mathrm{b}$ was constrained using the optical minimum variability timescale $\mathrm{log}(\Delta t_\mathrm{min})=3.75$~\citep{xiang2007}, where $\Delta t_\mathrm{min}$ is in units of seconds.
The angle between the line of sight of the observer and the jet axis, represented by $\theta_\mathrm{obs}$, was chosen to be equal to 1/$\Gamma$. 
This is referred to as the critical or superluminal angle, for which the Doppler factor equals $\Gamma$. 
The synchrotron emission was assumed to originate from relativistic electrons with Lorentz factors distributed between $\gamma_\mathrm{min}$ and $\gamma_\mathrm{max}$, following a power law with a spectral index $q_\mathrm{e}$, under the influence of a magnetic field $B$.
The particle-escape timescale is represented by $t_\mathrm{esc}$ = $\eta_\mathrm{esc}R_\mathrm{b}/c$, where $\eta_\mathrm{esc}$ is the particle-escape parameter. 
For each model, the parameters were adjusted to describe the data and achieve an equilibrium between the acceleration of the injected particles, the radiative cooling and the particle escape.
The best-fit parameters were used to calculate the relative partition between the magnetic field energy density and the kinetic luminosity of relativistic particles ($\epsilon_\mathrm{Be,p} \equiv L_\mathrm{B}/L_\mathrm{e,p}$) for each model.
All model spectra were corrected for extragalactic background light (EBL) absorption using the model of \cite{finke10}.
For the optical band, a phenomenological SED reproducing the archival host galaxy spectral points was added to the model.

The first model we applied assumed a pure SSC scenario. 
The magnetic field energy density required in this model is only 6\% of the value corresponding to equipartition with the relativistic electron distribution ($\epsilon_\mathrm{Be}  = 0.06$). 
The model spectrum is too hard in the \emph{Fermi} band (strongly curved, with $\Gamma\sim1.5$ around $10^{23}$\,Hz) and too soft in the VERITAS band ($\Gamma=4.0$), albeit within the errors in both cases.
On the other hand, while the X-ray measurements are well reproduced, the optical ($R$-band) spectrum is not. 
\par
Next, we tested a combined SSC+EC model. 
The external source of photons was assumed to be an isotropic thermal blackbody (BB) radiation field, which may be due to a torus of warm dust with temperature $T_\mathrm{ext}=1.5\times10^{3}$ $\textrm{K}$. 
The assumed BB infrared (IR) radiation field corresponds to a $\nu\mathrm{F_{\nu}}$ flux of $\sim 5\times10^{8}\times \mathrm{R}^{2}\,_{pc}$ JyHz, where $\mathrm{R}_\mathrm{pc}$ is the characteristic size of the IR emitter in units of parsecs. 
It should be noted that this quantity is far below the measured IR flux, thus consistent with our observations.
The addition of an EC component improves the modeling for the optical and \emph{Fermi} data compared with the pure SSC model and leads to values for the model parameters which are very close to equipartition ($\epsilon_\mathrm{Be}  = 1.20$).
However, the model tends to have too sharp a cutoff in the VHE band and therefore underpredicts the VERITAS flux measurements.
This could be remedied by choosing a much weaker magnetic field and higher electron energies, but the resulting system would then be very far from equipartition, with $\epsilon_\mathrm{Be}$ reduced by at least two orders of magnitude.
\par
The last model we tested is a combined lepto-hadronic jet model as described in \cite{boettcher2010}. 
In this case, the HE component of the non-thermal emission is dominated by a combination of synchrotron radiation from ultrarelativistic protons ($E_\mathrm{max}\gtrsim10^{19}$\,eV) and photons from decay of neutral pions. 
Secondary electrons that are produced in various electromagnetic cascades are the origin of the low-energy synchrotron emission. 
The kinetic energy of the relativistic proton population was assumed to have a single power-law distribution in the energy range $1.0\times10^{3}$\,GeV $< E_\mathrm{p} <$ $1.6\times10^{10}$\,GeV, with a spectral index $q_\mathrm{p}=2.4$. 
The model is a good description of the overall SED, and the system is close to equipartition between the magnetic field and the total relativistic particle content dominated by protons ($\epsilon_\mathrm{Bp}  = 0.95$). 
As is typical for lepto-hadronic models, the acceleration of protons to ultrarelativistic energies ($\sim10^{10}$GeV) requires a high magnetic field, 30\,G in this case. 
Although the lepto-hadronic model provides the best description for the data, it has two more free parameters than the SSC+EC model and is therefore less constraining. 
The best-fit parameters adopted for all three models are summarized in Table~\ref{tab:modeling}.

\begin{table}
\centering
\begin{small}
\begin{tabular}{|c|cccc|}
\hline\hline
Parameter & Symbol & SSC & SSC+EC & Lepto-hadronic\\
\hline
Electron low-energy cutoff*  				& $\gamma_\mathrm{min}$& $7.0\times10^{4}$  & $5.0\times10^{4}$ & $4.5\times10^{3}$ \\

Electron high-energy cutoff*  				& $\gamma_\mathrm{max}$& $1.0\times10^{6}$  & $1.0\times10^{6}$ & $5.0\times10^{4}$ \\

Injection electron spectral index*  			& $q_{e}$	& 2.4  		& 2.5 		& 2.4 \\

Escape-time parameter ($t_\mathrm{esc}$ = $\eta_\mathrm{esc}R_\mathrm{b}/c$)* 	& $\eta_\mathrm{esc}$ 	& 10  	& 300 		& 100 \\

Magnetic field (gauss)*					& $B$	  	& 0.1   	& 0.22  	& 30.0  \\

Bulk Lorentz factor* 					& $\Gamma$	& 20 		& 20 		& 15 	\\

Doppler factor						& D		& 20		& 20		& 15	\\

Blob radius ($\times10^{16}$$\textrm{cm}$)*		& $R_\mathrm{b}$ 	& 1.1 		& 1.6 		& 0.5 	\\

Observing angle 					& $\theta_\mathrm{obs}$& $2^{\circ}.87$& $2^{\circ}.87$ & $3^{\circ}.82$\\

External radiation field $E$ density ($\textrm{erg}$~$\textrm{cm}^{-3}$)*		& $u_\mathrm{ext}$	& .\,.\,.	& $6\times10^{-7}$ & .\,.\,.	\\

External radiation field BB temperature*		& $T_\mathrm{ext}$	& .\,.\,.		& $1.5\times10^{3}$ K	& .\,.\,.  	\\

Proton spectrum low-energy cutoff (GeV)*		& $E_\mathrm{p,min}$	& .\,.\,.		& .\,.\,. 		& $1.0\times10^{3}$	\\

Proton spectrum high-energy cutoff (GeV)*		& $E_\mathrm{p,max}$	& .\,.\,.		& .\,.\,. 		& $1.6\times10^{10}$	\\

Spectral index of proton distribution*			& $q_\mathrm{p}$	& .\,.\,. 		& .\,.\,. 		& 2.4			\\

Kinetic luminosity in protons ($\textrm{erg}$~$\textrm{s}^{-1}$)*& $L_\mathrm{p}$	& .\,.\,. 	& .\,.\,.		& $2.0\times10^{46}$	\\

Kinetic luminosity in electrons ($\textrm{erg}$~$\textrm{s}^{-1}$)*& $L_\mathrm{e}$(jet)	& $2.97\times10^{43}$ & $1.55\times10^{43}$ & $6.26\times10^{40}$	\\ 

Magnetic field energy density ($\textrm{erg}$~$\textrm{s}^{-1}$)& $L_\mathrm{B}$(jet) & $1.82\times10^{42}$ &	$1.86\times10^{43}$ & $1.90\times10^{46}$	\\

Equipartition parameter					& $\epsilon$	& $\epsilon_\mathrm{Be}  = 0.06$ & $\epsilon_\mathrm{Be}  = 1.20$ & $\epsilon_\mathrm{Bp}  = 0.95$ \\ 

Redshift 						& $z$		& 0.19 & 0.19 & 0.19 \\
\hline\hline
\end{tabular}
\end{small}
\caption{{\small Best-fit parameters for the SED of RBS\,0413 for the SSC, SSC+EC, and lepto-hadronic models. The parameters that were left free are marked with an asterisk. $\theta_\mathrm{obs}$ is the superluminal angle (see Section~\ref{modeling}).}}
\label{tab:modeling}
\end{table}

Based on our calculations, all three models are good at describing the observed data. 
It appears that if the criterion of equipartition is taken as a reasonable measure of successful blazar emission models, SSC+EC is preferred over SSC for this HBL, which seems to be in contrast with some previous blazar studies. 
See~\citet{ghisellini98} for arguments relating the presence of an EC component with the blazar sequence and~\citet{abdo2010b} for a discussion of issues encountered in explaining blazar SEDs with a simple one-zone homogeneous SSC model.
On the other hand, we cannot discriminate between leptonic and lepto-hadronic mechanisms, since the SSC+EC and lepto-hadronic models provide equally reasonable descriptions for the observed non-thermal continuum, and we did not detect variability in the HE and VHE regimes given the limited statistics.  
Since the synchrotron cooling timescales for electrons and protons are different, the detection of intraday variability would be harder to explain with a lepto-hadronic scenario and would accordingly favor a purely leptonic scenario. 
Therefore, any future observation of rapid variability would be helpful in distinguishing between the SSC+EC and lepto-hadronic models. 

\acknowledgements
The VERITAS research is supported by grants from the US Department of Energy Office of Science, the US National Science Foundation, the
Smithsonian Institution, and the NASA \textit{Swift} Guest Investigator Program, by NSERC in Canada, by Science Foundation Ireland (SFI 10/RFP/AST2748), and by STFC in the UK. We acknowledge the
excellent work of the technical support staff at the FLWO and at the collaborating institutions in the construction and
operation of the instrument.

The \emph{Fermi}-LAT Collaboration acknowledges generous ongoing support from a number of agencies and institutes that have supported both the
development and the operation of the LAT as well as scientific data analysis.
These include the National Aeronautics and Space Administration and the Department of Energy in the United States, the Commissariat \`a l'Energie Atomique
and the Centre National de la Recherche Scientifique / Institut National de Physique Nucl\'eaire et de Physique des Particules in France, the Agenzia Spaziale Italiana and the Istituto Nazionale di Fisica Nucleare in Italy, the Ministry of Education, Culture, Sports, Science and Technology (MEXT), High Energy Accelerator Research Organization (KEK) and Japan Aerospace Exploration Agency (JAXA) in Japan, and the K.~A.~Wallenberg Foundation, the Swedish Research Council and the Swedish National Space Board in Sweden. Additional support for science analysis during the operations phase is gratefully acknowledged from the Istituto Nazionale di Astrofisica in Italy and the Centre National d'Etudes Spatiales in France.

{\it Facilities:} \facility{VERITAS}, \facility{\textit{Fermi}}.


\bibliographystyle{apj}
\begin{multicols}{2}

\end{multicols}
\end{document}